\documentclass[debug,overfull,cite,figures]{epl}%
\usepackage{graphicx}
\usepackage{amsmath}
\usepackage{amsfonts}%
\usepackage{amssymb}

\begin{document}

\title{Scattering induced dynamical entanglement \\ and the quantum-classical
correspondence}
\shorttitle{Scattering induced dynamical entanglement}
\author{M. Lombardi \and A. Matzkin}
\institute{Laboratoire de Spectrom\'{e}trie physique
(CNRS Unit\'{e} 5588), Universit\'{e} Joseph-Fourier Grenoble-1,
BP 87, 38402 Saint-Martin, France}

%\pacs {03.65.Ud}{Entanglement and quantum nonlocality}

 \maketitle

\begin{abstract}
The generation of entanglement produced by a local potential interaction in a
bipartite system is investigated. The degree of entanglement is contrasted with
the underlying classical dynamics for a Rydberg molecule (a charged particle
colliding on a kicked top). Entanglement is seen to depend on the structure of
classical phase-space rather than on the global dynamical regime. As a
consequence regular classical dynamics can in certain circumstances be
associated with higher entanglement generation than chaotic dynamics. In
addition quantum effects also come into play: for example partial revivals,
which are expected to persist in the semiclassical limit, affect the long time
behaviour of the reduced linear entropy. These results suggest that
entanglement may not be a pertinent universal signature of chaos.

\end{abstract}

Entanglement, i.e. the nonseparability intrinsic to composite
systems, is one of the most peculiar features of the quantum world.
In its most popular form, found for example in the original EPR
proposal \cite{EPR1935}, the entanglement is of geometrical nature.
However any usual potential interaction between two particles can
generically lead to entanglement, provided several quantum states
are accessible to both particles. The study of such dynamical
entanglement is particularly interesting for systems which possess a
classical counterpart. Indeed as is well-known \cite{spbook}, such
systems can be investigated with semiclassical tools, allowing to
interpret the quantum dynamics in terms of classical properties.
Recent investigations have been focusing on the relation between the
generation of entanglement and the underlying classical dynamics.
Initial work in spin-boson systems \cite{furuyaetal98} and in
coupled kicked tops \cite{millersarkar99} suggested that chaotic
dynamics generate more and faster entanglement. This claim was
subsequently revised \cite{fu03,aguiar04} and efforts are now being
made to derive universal relations ruling the generation of
entanglement irrespective of system specifics. Different approaches
based on perturbation expansions \cite{fu03,kus04,furuya05}, Random
Matrix theory \cite{seligmangorin03,laksha04} and semiclassical
methods \cite{jacquod04,prosen05} have been proposed. In particular
it was shown and verified \cite{jacquod04,jacquod05} that
entanglement generated from initial Gaussian states averaged over
configuration space depends on the global classical dynamical
regime. However contrarily to the well established quantum-classical
correspondence for spectral statistics and large scale fluctuations
\cite{haake}, it is still not clear to what extent an analog
universal correspondence exists for dynamical entanglement
production; for example the dependence of entanglement on non
classical features like the type of initial states \cite{kus04} or
their spectral width \cite{seligmangorin03} has been put forward.

In this work we study the generation of entanglement for a real system which
possesses an unambiguous classical counterpart. This will be done by employing
a bipartite system where entanglement is produced by a \emph{local}
interaction, viz. the particles interact via a short-range scattering
potential. Our model enables more stringent tests on the transition
to chaos than usual models in which this transition depends essentially
on a single coupling parameter and occurs relatively uniformly on the whole
phase space. Indeed we use a system in which a second essential parameter, the
resonance parameter, produces global changes of the phase space picture,
changing dramatically the coupling necessary to induce chaos.  We will see
that the underlying classical dynamics is reflected in the generation of
entanglement. However this dependence involves particular features in the
structure of phase-space rather than the global dynamical (chaotic or regular)
regimes. Quantum effects also come into play, particularly at long times.

\begin{figure}[tb]
\fourimages[angle=270,width=0.23\textwidth]{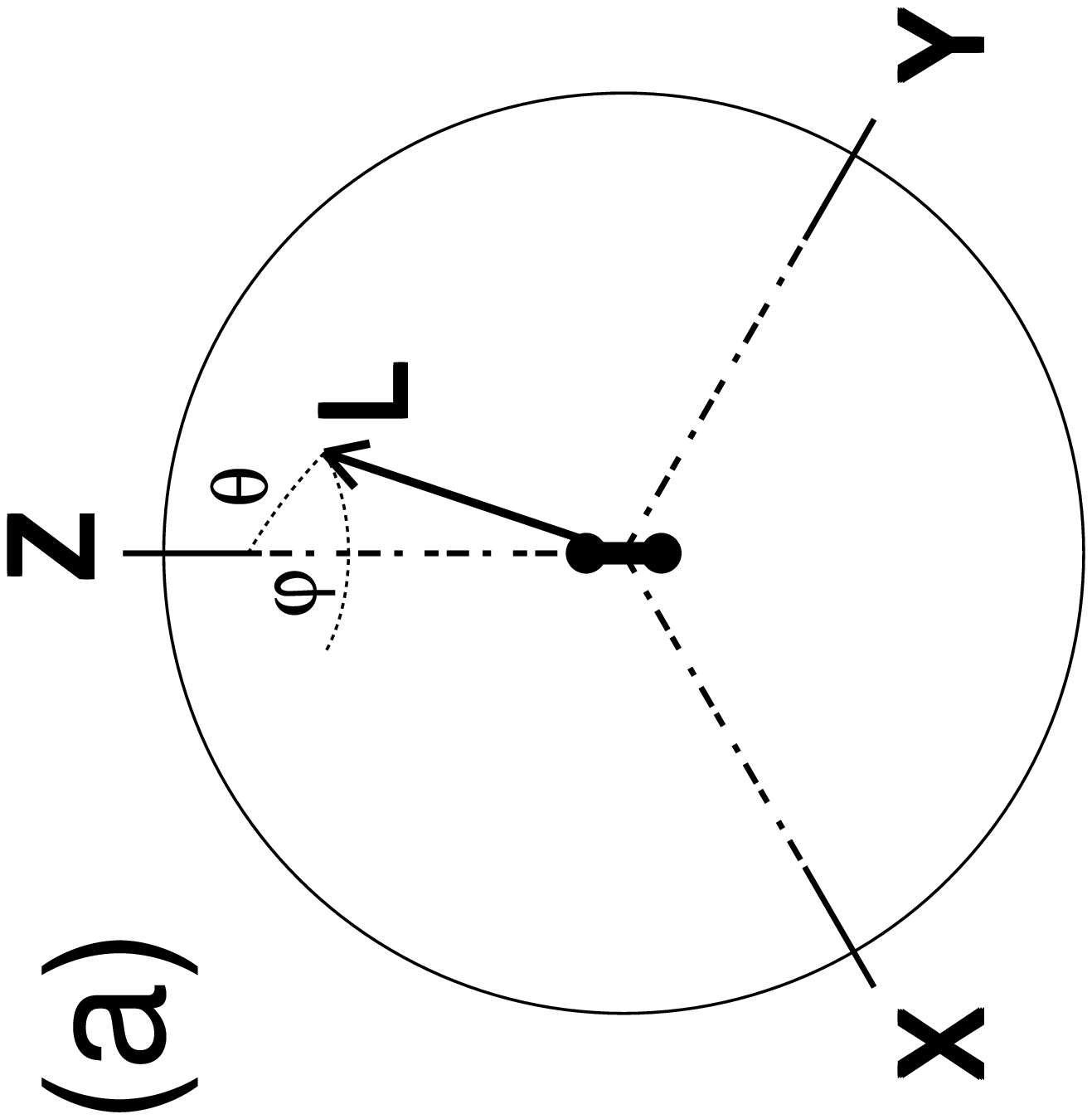}{fig1b}{fig1c}{fig1d}
\fourimages[angle=270,width=0.23\textwidth]{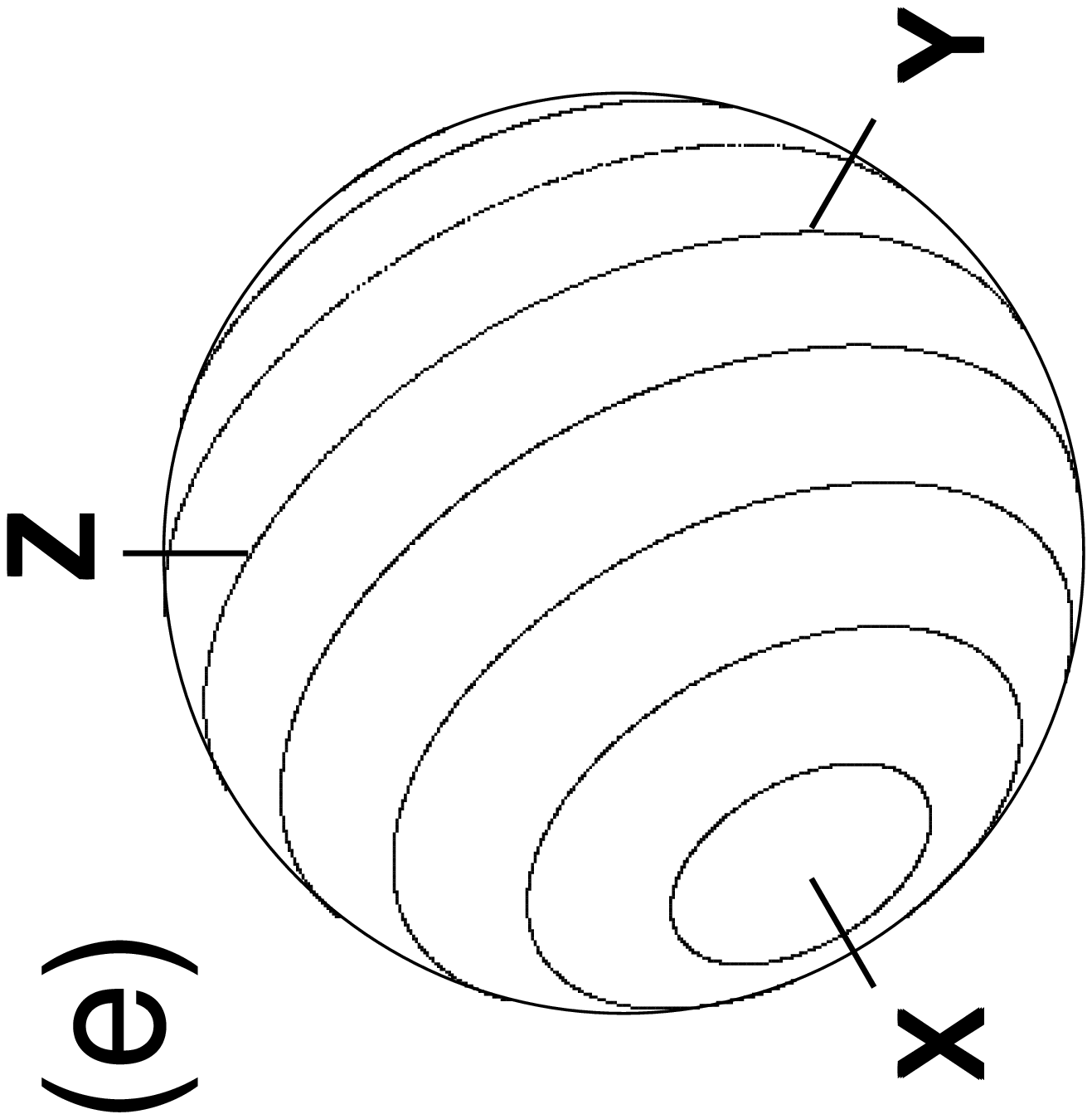}{fig1f}{fig1g}{fig1h}
\caption{Poincar\'e surfaces of section in the molecular frame.
(a) defines the angles $\theta$ and $\varphi$ for a given position
of $\vect{L}$ on the sphere. The core axis is along $OZ$ and
$\vect{N}$ along $OX$. (e) shows the k=0, no coupling situation,
valid for all values of $T_{e}$ and $T_{0}$; one sees only the
free apparent rotation of $\vect{L}$ around the core angular
momentum $\vect{N}$, describing circles. On a given circle, the
modulus $N$ is constant, but it varies from circle to circle.
(b)-(d) and (f)-(h): generic case (top row) and resonant ($T_{e} =
T_{0}$) case (bottom row) for increasing coupling. (b) and (f):
$k=0.25$, (c) and (g): $k=0.5$, (d) and (h): $k=10$. Note that in
the generic case, circles (torii in whole phase space) which
correspond to low rational values of the returning times of
electron and core open resonant islands: values 2 and 4 appear
clearly in (b). Chaos appears along the corresponding unstable
manifolds and invades rapidly the whole phase space until the last
islands of regularity disappear [(d)]. In the resonant case a very
large couple of resonance islands appears around the $\pm OZ$ axis
(the top one is visible in (f)-(h)), with a separatrix starting
from the $\pm OY$ axis.} \label{fig1}
\end{figure}

The system investigated here is a Rydberg molecule. Simply stated,
a Rydberg molecule is composed of two parts: a highly excited
electron on the one hand, and a compact ionic molecular core,
containing the nuclei and the tightly bound other electrons on the
other. Most of the time, the outer electron and the core are
spatially well-separated: the core rotates freely with angular
momentum $\vect{N}$ and energy $E_{N}^{+}\varpropto N^{2}$,
whereas the outer electron senses a pure Coulomb field. Its
orbital angular momentum $\vect{L}$ is fixed in space. Seen in the
molecular frame $\vect{L}$ turns around the $\vect{N}$ axis (see
fig.~\ref{fig1}(a)). However, the outer electron periodically
scatters on the molecular core. Classically the electron is kicked
by the rotating core. This kick results in a change in the
direction of $\vect{L}$ by $\Delta\varphi$, the deflection angle
of the plane of the classical Kepler orbit. $\vect{N}$ adjusts
accordingly, since the total angular momentum $\vect{J} = \vect{L}
+ \vect{N}$ is conserved. The conservation of $\vect{L}$ and the
cylindrical symmetry of the core implies that $\Delta\varphi =
k\cos\theta$ where $k$ gives the strength of the kick. A
negligibly small kick yields regular dynamics. This is visualised
in a Poincar\'{e} surface of section obtained by plotting the
position of $\vect{L}$ after each kick (fig.~\ref{fig1})
\cite{lombardi}. As $k$ increases, two cases must be
distinguished. In the generic case [Figs.~\ref{fig1}(b)-(d)] chaos
appears and invades rapidly the whole phase space until the last
islands of regularity disappear. Things are different in the
resonant case. Resonances appear when the orbital period of the
electron $T_{e}$ and the rotational period of the core $T_{0}$ are
commensurate, so resonances are nongeneric. Chaos is inhibited and
finally appears along the separatrix [Figs.~\ref{fig1}(f)-(h)] for
much higher values of the coupling $k$. Such resonances are known
to modify the spectrum of small Rydberg molecules, giving rise to
the stroboscopic effect experimentally observed on Na$_{2}$
\cite{exp}.

In quantum-mechanical terms the collision couples the electron and core
dynamics and induces phase-shifts in the wavefunction of the outer electron. In
the absence of the core-electron interaction, the Hamiltonian $H_{0}$ is
separable and the eigenstates are given by the product state
\begin{equation}
\phi_{N}(E,r)=f_{L}(E-E_{N}^{+},r)\otimes\left|  N\right\rangle \label{5}
\end{equation}
where $E$ is the total energy, $\left|  N\right\rangle $ the state of the core
and $f_{L}$ the radial Coulomb function of the electron whose energy is
$\epsilon_{N}=E-E_{N}^{+}$ \cite{note1}. The total Hamiltonian $H=H_{0}+V$
includes the core-electron interaction potential $V$. The eigenstates are given
from scattering theory by%
\begin{equation}
\psi(E,r)=\sum_{N}Z_{N}(E)\left[  I+G_{0}(E)K\right]  \phi_{N}(E,r),
\label{10}
\end{equation}
where $G_{0}$ is the Green's function, $Z_{N}$ are expansion coefficients and
$K$ is the scattering matrix, whose elements $K_{N^{\prime}N}$ are related to
the transition probability amplitude between states $\phi_{N}$ and
$\phi_{N^{\prime}}$ due to the collision. From eq.~(\ref{10}) it is apparent
that an eigenstate is given by a superposition of core states with rotational
number $N$, each core state being associated with an outer electron having an
energy $\epsilon_{N}$. The relation between the quantum and classical pictures
is simpler \cite{lombardi} when expressed in the molecular frame, where $K$ has
only diagonal elements $\tan\delta_{\Lambda}$. The phase-shifts
$\delta_{\Lambda}$ only depend on the coupling between $\vect{L}$ and the
molecular axis. $\Lambda=L\cos\theta$ is the projection of $\vect{L}$ on this
axis and the classical deflection angle is related to the quantum phase shifts
by
\begin{equation}
\Delta\varphi = 2 \frac{\partial\delta_{\Lambda}}{\partial\Lambda} = k
\frac{\Lambda}{L}.
\end{equation}
The first equality is a very general result stemming from the semiclassical
approximation to scattering theory \cite{newton} and is valid to first order in
$\hbar$, whereas the last term follows from the functional relation
$\delta_{\Lambda}(k)$ chosen in this work which turns out to be verified for
typical diatomic molecules.

\begin{figure}[tb]
\onefigure[width=2.25in]{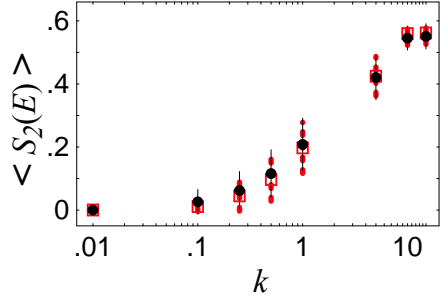} \caption{Average linear entropy
for generic (red/gray boxes) and resonant (black dots) states
($T_{e}=T_{0}$), shown for different values of $k$. The rms is also
shown as red dashed (generic) and solid (resonant) error bars.}
\label{fig2}%
\end{figure}

To quantify entanglement we will employ the linear entropy $S_{2}$,
as has been done in most previous works
\cite{furuyaetal98,fu03,aguiar04,kus04,furuya05,seligmangorin03,laksha04,jacquod04,prosen05}.
Indeed for bipartite systems $S_{2}$ behaves as other measures of
entanglement \cite{BS03} while being convenient to compute.
 The linear entropy associated with  the reduced density matrix describing the electron
 is given by
\begin{equation}
S_{2}=1-\mathrm{Tr}_{e}\rho_{e}^{2}, \label{20}
\end{equation}
where $\rho_{e}=\mathrm{Tr}_{c}\rho$ and as usual $\mathrm{Tr}_{c}$
($\mathrm{Tr}_{e}$) refers to averaging over the core (electron) degrees of
freedom. We first determine the degree of entanglement of stationary states.
This is done by calculating $\rho(E)=\left|  \psi(E)\right\rangle \left\langle
\psi(E)\right|  $ over an energy range for which the classical dynamics does
not vary appreciably. The values $L=2$ and $J=10$ adopted in this work
correspond to experimentally accessible values for diatomic molecules such as
Na$_{2}$. With these values and taking into account symmetry requirements,
three rotational states are accessible to the core $N=8,10,12$ (classically,
$8\leq N\leq12$). Hence in a typical eigenstate (\ref{10}) three core-electron
states are entangled. The mean entropy $\left\langle S_{2}(E)\right\rangle $
for stationary states is shown in fig.~\ref{fig2} as a function of the coupling
constant $k$. The curves for the generic and resonant cases are very close
despite phase-space being substantially different for most values of $k$. On
average the potential interaction entangles at least as efficiently when the
underlying classical dynamics is mostly regular rather than mainly chaotic.
Note the rms is small for the regular and totally chaotic regimes and larger in
the mixed phase-space situation and that the linear entropy saturates in both
the generic and resonant cases only for large $k$: $\left\langle
S_{2}(E)\right\rangle $ approaches its upper limit $2/3$ corresponding to a
maximally mixed state.

We next determine the time-dependent generation of entanglement from an initial
product state. At $t=0$ we take the wavefunction of the electron to be radially
localised at the outer turning point $r_{tp}$, several thousand atomic units
away from the core, and we take the core to be in the rotational state $\left|
N_{0}\right\rangle $, so that
\begin{equation}
\psi(t=0,r)=F_{loc}(r\approx r_{tp})\otimes\left|  N_{0}\right\rangle .
\end{equation}
The mean energy of the radial wavepacket $F_{loc}$ is $\epsilon_{0} =
E_{0}-E_{N_{0}}^{+}$ and $T_{e}$ is the period of the corresponding Kepler
orbit. At $t\approx T_{e}/2$ the wavepacket scatters off the core and the
wavefunction is given by a \emph{superposition} of radial wavepackets moving
outward, each associated with a core in a given rotational state $\left|
N\right\rangle $. At $t\approx T_{e}$ each of the radial wavepackets has
reached the outer turning point and starts going back towards the core. The
core-electron interaction then takes place again at $t\approx3T_{e}/2$
resulting in further entanglement change. As $t$ increases the wavepackets tend
to spread radially resulting in a continuous core-electron interaction. These
features appear in the time dependence of the linear entropy $S_{2}(t)$,
obtained from eq.~(\ref{20}) by tracing out the core degrees of freedom from
the total density matrix $\left|  \psi(t)\right\rangle \left\langle
\psi(t)\right|  $.

\begin{figure}[tb]
\onefigure[width=2.25in]{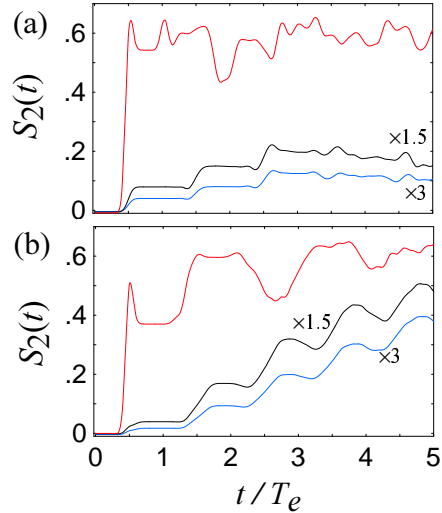} \caption{Short time variation of
the linear entropy for different values of the coupling strength;
$t$ is given in units of the Kepler period $T_{e}$. (a) Generic
phase-space, from top to bottom $k=10$ (red), 0.5 (black) and 0.25
(blue). The $k=0.25$ ($k=0.5$) curve is multiplied by a factor 3
(1.5). (b) Same as (a) for the resonant case $T_{e}=T_{0}$.}
\label{fig3}
\end{figure}

For short times $S_{2}(t)$ is shown in fig.~\ref{fig3}, for $N_{0}=8$ and for 3
different values of the coupling $k$. In the generic case, for small $t$ and
not too large $k$, $S_{2}$ is seen to increase linearly in steps, reflecting
the discrete times core-electron interaction, and then appears to saturate (but
see below). In the resonant case, the linear entropy also increases on average
linearly but displays oscillations. Indeed, a semiclassical approximation for
the short-time behaviour of the purity $\mathrm{Tr}_{e} \rho_{e}^{2}$ predicts
that $S_{2}(t_{C})\propto k^{2}t_{C}$. This result, valid only for the
collision times $t_{C}=(2m+1)T_{e}/2$, is obtained in the stationary phase
approximation by neglecting correlations between electron orbits built on
different core rotational states. This linear behaviour agrees with the
prediction of \cite{prosen05}. The $k^{2}$ dependence comes from the
semiclassical amplitudes, and the linear variation in $t$ from the fact that
these amplitudes are conserved.
The oscillations of $S_{2}$ observed in the resonant case arise from the
coherent projection of the motion of the outer electron. It is in this sense a
kinematical effect, since no such behaviour is to be found in the wavefunction
or in the probability flux. By expanding the core energies to first order, the
purity is seen to depend on terms of the form
\begin{equation}
\sum_{NN^{\prime}} | e^{-2i\pi t\left[  (N-N^{\prime})/T_{0}\right] }
\sum_{\epsilon(N)\epsilon^{\prime}(N^{\prime})}e^{-i(\epsilon-\epsilon
^{\prime})t}Z_{N}(E)Z_{N^{\prime}}(E^{\prime})\left\langle F_{L}
(\epsilon^{\prime}(N^{\prime}))\right|  \left.  F_{L}(\epsilon
(N))\right\rangle |^2,\label{25}%
\end{equation}
where $\epsilon_N=E-E_{N}^{+}$ and $F_{L}(\epsilon,r)$ are radial functions on
the energy shell (i.e. a combination \ of regular and irregular Coulomb
functions with effective phase shifts depending on the scattering matrix $K$).
These functions form an overcomplete basis and are not orthogonal in
$\epsilon$. In the generic case the\ nondiagonal terms in the second sum are
generally small and uncorrelated. However at resonance both exponentials are in
phase (since $\epsilon-\epsilon^{\prime}\approx 2m\pi/T_{0}$) and the
correlations between nondiagonal terms are enhanced, because the resonance
imposes a greater similarity of certain corresponding eigenstates (\ref{10})
\cite{lombardi}.

\begin{figure}[tb]
\onefigure[width=2.25in]{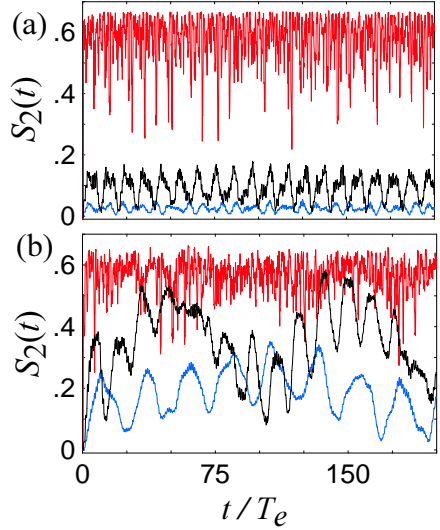} \caption{Same as fig. 3 for
longer times; the curves for $k=0.25$ and $k=0.5$ are \emph{not}
multiplied by any factor. The oscillations visible for these curves
depend on the revival time of the partial correlation functions. For
example in (b) for $k=0.5$, two  scales are visible corresponding to
$T_{e}^{rev}(N=10)$ and the longer $T_{e}^{rev}(N=12)$.}
\label{fig4}%
\end{figure}

The long time behaviour of the entanglement production is shown in
fig.~\ref{fig4}. In the generic case, $S_{2}(t)$ is on average very
low for $k=0.25$ (corresponding to classically regular phase space),
higher for $k=0.5$ (mixed phase-space) and nearly maximal for $k=10$
(chaotic phase space). Periodic recoherences are visible for small
$k$. Although revivals are expected when phase space is regular and
the dimensionality of the Hilbert space is low, we point out that
the same observations have been made for considerably larger angular
momenta closer to the semiclassical limit \cite{ma-prep}. The
characteristic periods of recoherence involve the revival times
$T_{0}^{rev}$ and $T_{e}^{rev}$ of the core rotation and the
electron orbit. The former readily appears by carrying the expansion
in $E_{N}^{+}$ in eq.~(\ref{25}) to second order. The latter follows
by remarking that the second sum in eq.~(\ref{25}) is the
cross-correlation function between parts of the wavefunction
belonging to different superposition alternatives (i.e. different
scattering channels). Alternatively, it can be shown
\cite{prosenetal03} that the purity is bounded from below by
correlation functions which vary according to the revival times.
Note that the revivals have no classical counterpart. In the
resonant case the long-range oscillations can be dramatically
enhanced by the commensurate time scales (in addition to
$T_{0}=T_{e}$, we have $T_{0}^{rev}\approx2T_{e}^{rev}$) and the
similarity of the eigenfunctions which makes the cross-correlations
to be very large. The consequence is that the curves for $k$ values
corresponding to different underlying classical dynamics may cross.

These results indicate that the behaviour of the linear entropy is
not simply connected to the classical regime, but also depends on
the structure of phase-space. Indeed the entanglement production is
considerably \textit{higher} in the resonant case despite classical
phase space being much more \textit{regular} (compare
Figs.~\ref{fig1}(c) and (g) with their respective curves in
fig.~\ref{fig4}). This behaviour can be ascribed to the large values
of the cross correlations, which are ultimately due to the fact that
at resonance $\vect{L}$ appears to move around the core axis $OZ$,
and therefore its projection on $\vect{N}=OX$ (and thus the possible
values of $N^{2}$) tend to spread. Classically, resonances modify
the structure of phase-space by the appearance of a fixed point on
the $OZ$ core axis, which is visible in the Poincar\'{e} surfaces of
section (fig.~\ref{fig1}). In quantum mechanical terms $N$ is then
not well-defined and thus the wavefunctions are mixed in $\left|
N\right\rangle $ even for small kick strengths. In the generic case
however these mixings can only take place for large values of the
coupling, corresponding to chaotic dynamics.

To conclude we have investigated the generation of entanglement
induced by a potential interaction in a bipartite system, a Rydberg
molecule. By and large chaotic classical dynamics is associated with
more and faster entanglement because the scattering particles
explore larger portions of phase-space leading quantum-mechanically
to superpositions. But chaos is not necessarily the most efficient
way to ensure that the relevant part of phase-space leading to
superpositions (in our system the rate of inelastic scattering) is
explored with a large classical amplitude: this is why a dynamical
effect such as resonances generate higher entanglements despite
classical phase-space being mostly regular. On the other hand it has
been shown \cite{jacquod04,jacquod05} that for weak coupling the
purity does depend on the global dynamical regime when a given type
of initial state is averaged over all the possible initial
positions. But it is also known that the average entanglement is
sensitive to the form of the initial state \cite{kus04}. From a
general standpoint we would like to point out that the universal
relations between quantum fluctuations and average classical
phase-space properties are known to be spoiled by nongeneric
families of periodic orbits \cite{aj1} that are expected to
proliferate in systems (such as Rydberg systems) that are strongly
coupled by potential scattering \cite{mm04}  which is here the
generator of entanglement.

\acknowledgments We are grateful to T. H. Seligman (UNAM,
Cuernavaca, Mexico) for fruitful discussions.


\begin{thebibliography}{00}

\bibitem {EPR1935} \Name{ Einstein A., Podolsky B. \and Rosen N.}
                   \REVIEW{Phys. Rev.}{47}{1935}{777}

\bibitem {spbook} \Name{Brack M. \and Bhaduri R. K.}
                 \Book{Semiclassical Physics}
                 \Publ{Addison-Wesley, Reading, USA}
                 \Year{1997}

\bibitem {furuyaetal98} \Name{ Furuya K., Nemes M. C. \and Pellegrino G. Q.}
                   \REVIEW{Phys. Rev. Lett.}{80}{1998}{5524}

\bibitem {millersarkar99} \Name{Miller P. A. \and Sarkar S.}
                          \REVIEW{Phys. Rev. E}{60}{1999}{1542}

\bibitem {fu03} \Name{Fujisaki H., Miyadera T. \and Tanaka A.}
                \REVIEW{Phys. Rev. E}{67}{2003}{066201}

\bibitem {aguiar04} \Name{Novaes M. \and de Aguiar M. A. M.}
                    \REVIEW{Phys. Rev. E}{70}{2004}{045201(R)}

\bibitem {kus04} \Name{Demkowicz-Dobrzanski R. \and Kus M.}
                 \REVIEW{Phys. Rev. E}{70}{2004}{066216}

\bibitem {furuya05} \Name{Angelo R. M. \and Furuya K.}
                    \REVIEW{Phys. Rev. A}{71}{2005}{042321}

\bibitem {seligmangorin03} \Name{Gorin T. \and Seligman T. H.}
                           \REVIEW{Phys. Lett A}{309}{2003}{61}

\bibitem {laksha04} \Name{Bandyopadhyay J. N. \and Lakshminarayan A.}
                    \REVIEW{Phys. Rev. E}{69}{2004}{016201}

\bibitem {jacquod04} \Name{Jacquod Ph.}
                     \REVIEW{Phys. Rev. Lett.}{92}{2004}{150403}

\bibitem {prosen05} \Name{Znidaric M. \and Prosen T.}
                    \REVIEW{Phys. Rev. A}{71}{2005}{032103}

\bibitem {jacquod05} \Name{Petitjean C. \and Jacquod Ph.}
                     \REVIEW{Eprint}{\hspace{.01cm}}{2005}{quant-ph 0510157}

\bibitem {haake} \Name{Haake F.}
                 \Book{Quantum signatures of chaos}
                 \Publ{Springer, Berlin}
                 \Year{2001}

\bibitem {lombardi} \Name{Dietz B., Lombardi M. \and Seligman T. H.}
                    \REVIEW{Ann. Phys.} {312}{2004}{441};
                    \Name{Lombardi M. \etal}
                    \REVIEW{J. Chem. Phys.}{89}{1988}{3479}

\bibitem {exp} \Name{Labastie P. \etal}
               \REVIEW{Phys. Rev. Lett.}{52}{1984}{1681};
               \Name{Chang  E. S. \etal}
               \REVIEW{J. Chem. Phys.}{111}{1999}{6247}

\bibitem {note1} The angular part of the electron wavefunction is contained in
the compound notation $\left|  N\right\rangle $ due to the coupling of angular
momenta. This 'geometrical entanglement' plays no role in this work which
focuses on the superpositions in $\vect{N}^{2}$ and not in the relative
orientations of the angular momenta.

\bibitem {newton} \Name{Newton R. G.}
                  \Book{Scattering theory of waves and particles}
                  \Publ{Springer, New-York}
                  \Year{1982}
                  Ch. 18

\bibitem {ma-prep} \Name{Lombardi M. \and Matzkin  A.} in preparation.

\bibitem {BS03} \Name{Berry D W \and Sanders B C}
                        \REVIEW{J. Phys. A}{36}{2003}{12255}

\bibitem {prosenetal03} \Name{Prosen T., Seligman T. H. \and Znidaric M.}
                        \REVIEW{Phys. Rev. A}{67}{2003}{062108}

\bibitem {aj1} \Name{H\"{o}nig A. \and Wintgen D.}
               \REVIEW{Phys. Rev. A}{39}{1989}{5642};
               \Name{Sieber M. \etal}
               \REVIEW{J. Phys. A}{26}{1993}{6217}

\bibitem {mm04}\Name{Matzkin A. \and Monteiro T. S.}
               \REVIEW{J. Phys. A}{37}{2004}{L225}

\end{thebibliography}
\end{document}